\documentclass[prl,amsmath,amssymb,floatfix,superscriptaddress,notitlepage,showpacs,twocolumn]{revtex4}
\usepackage{graphicx}
\usepackage{hyperref}
\usepackage{bm}
\usepackage{amssymb}

\begin{document}

\title{Information-Entropic Signature of the Critical Point}

\author{Marcelo Gleiser}
\email{mgleiser@dartmouth.edu}
\affiliation{Department of Physics and Astronomy, Dartmouth College,
Hanover, NH 03755, USA}

\author{Damian Sowinski}
\email{Damian.Sowinski.GR@dartmouth.edu}
\affiliation{Department of Physics and Astronomy, Dartmouth College,
Hanover, NH 03755, USA}

\date{\today}

\begin{abstract}

We investigate the critical behavior of continuous (second-order) phase transitions in the context of (2+1)-dimensional Ginzburg-Landau models with a double-well effective potential. In particular, we show that the recently-proposed configurational entropy (CE)--a measure of the spatial complexity of the order parameter based on its Fourier-mode decomposition--can be used to  identify the critical point. We compute the CE for different temperatures and show that large spatial fluctuations near the critical point  ($T_c$)--characterized by a divergent correlation length--lead to a sharp decrease in the associated configurational entropy. We further show that the CE density has a marked scaling behavior near criticality, with the same power of Kolmogorov turbulence. We reproduce the behavior of the CE at criticality with a percolating many-bubble model.

\end{abstract}


\maketitle

\section{Introduction}
From materials science \cite{Materials} to the early universe \cite{CosmoPT}, phase transitions offer a striking illustration of how changing conditions can affect the physical properties of matter \cite{Landau}. In very broad terms, and for the simplest systems described by a single order parameter, it is customary to classify phase transitions as being either discontinuous or continuous, or as first or second order, respectively. First-order phase transitions can be described by an effective free-energy functional (or an effective potential in the language of field theory) where an energy barrier separates two or more phases available to the system. The system may transition from a higher to a lower free-energy state (or from a higher to a lower vacuum state) either by a thermal fluctuation of sufficient size (a critical bubble) or, for low-temperatures, by a quantum fluctuation. Generally speaking, this description of thermal bubble nucleation is valid as long as ${\cal F}[\phi_b]/k_BT \gg 1$, where ${\cal F}[\phi_b]$ is the 3d Euclidean action of the spherically-symmetric critical bubble or bounce $\phi_b(r)$, $k_B$ is Boltzmann's constant, and $T$ is the environmental temperature. For quantum tunneling, one uses instead ${\cal S}_4[\phi_b]/\hbar$, where ${\cal S}_4[\phi_b]$ is the $O(4)$-symmetric Euclidean action of the 4d bounce.

For second-order transitions the order parameter varies continuously as an external parameter such as the temperature is changed \cite{Landau}. A well-known example is that of an Ising ferromagnet, where the net magnetization of a sample is zero above a critical temperature $T_c$, the Curie point, and non-zero below it. Below $T_c$ the transition unfolds via spinodal decomposition, whereby long-wavelength fluctuations become exponentially-unstable to growth. This growth is characterized by the appearance of domains with the same net magnetization which compete for dominance with their neighbors. In the continuum limit, systems in the Ising universality class can be modeled by a Ginzburg-Landau (GL) free-energy functional with an order parameter $\phi({\bf x})$ \cite{Goldenfeld}. In the absence of an external source (or a magnetic field), the GL free-energy functional is simply
\begin{equation}
E[\phi] = \int d^dx \left [\frac{\gamma}{2}(\nabla\phi)^2 +\frac{a}{2} t \phi^2 + \frac{b}{4}\phi^4\right ],
\label{GL}
\end{equation}
\noindent
where $t\equiv (T-T_c)$, and $\gamma$, $a$, $b$ are positive constants. For $T > T_c$ the system has a single free-energy minimum at $\phi=0$, while for $T < T_c$ there are two degenerate minima at $\phi_0 = \pm (-at/b)^{1/2}$. This mean-field theory description works well away from the critical point. In the neighborhood of $T_c$ one uses perturbation theory and the renormalization group to account for the divergent behavior of the system. This behavior can be seen through the two-point correlation function $G({\bf r})$: away from $T_c$ $G({\bf r})$ behaves as $\sim \exp[-{\bf r}/\xi(T)]$, where $\xi(T)$ is the correlation length, a measure of the spatial extent of correlated fluctuations of the order parameter. In mean-field theory, $\xi(T)\sim |T-T_c|^{-\nu}$, where $\nu=1/2$, independently of spatial dimensionality. In the neighborhood of $T_c$, where the mean-field description breaks down, the behavior of spatial fluctuations is corrected using the renormalization group. Within the $\varepsilon$-expansion, one obtains, in 3d, $\nu = 1/2+\varepsilon/12 +7\varepsilon^2/162 \simeq 0.63$ \cite{Goldenfeld}.

For continuous transitions in GL-systems, the focus of the present letter, the critical point is characterized by having fluctuations on all spatial scales. This means that while away from $T_c$ large spatial fluctuations are suppressed, near $T_c$ they dominate over smaller ones. In this letter, we will explore this fact to obtain a new measure of the critical point based on the information content carried by fluctuations at different momentum scales. For this purpose we will use a measure of spatial complexity known as configurational entropy (CE), recently proposed by Gleiser and Stamatopoulos \cite{GS1}. The CE has been used to characterize the information content \cite {GS1} and stability of solitonic solutions of field theories \cite{GS1,GSo}, to obtain the Chandrasekhar limit of compact Newtonian stars \cite{GSo}, the emergence of nonperturbative configurations in the context of spontaneous symmetry breaking \cite{GS2} and in post-inflationary reheating \cite{InfRev} in the context of a top-hill model of inflation \cite{GG}. (Note that our usage of the name ``configurational entropy'' differs from other instances in the literature, as for example in protein folding \cite{Protein}.) Here we show that in the context of continuous phase transitions the CE provides a very precise signature and a marked scaling behavior at criticality.

\section{Model and Numerical Implementation}
We consider a (2+1)-dimensional GL model where the system is in contact with an ideal thermal bath at temperature $T$. The role of the bath is to drive the system into thermal equilibrium. This can be simulated as a temperature-independent GL functional (so, setting $t=-1$ in Eq. \ref{GL}) with a stochastic Langevin equation
\begin{equation}
\ddot{X}+\eta \dot X - \nabla^2 X-X+X^3+\xi=0,
\label{Langevin}
\end{equation}
\noindent
where we have introduced dimensionless variables $x^\mu=\frac{1}{\sqrt{a}}y^\mu$ and $\phi=\sqrt\frac{a}{b}X$, and we took $\gamma=1$. In Eq. \ref{Langevin}, $\eta$ is the viscosity and $\xi({\bf x},t)$ is the stochastic driving noise of zero mean, $\langle \xi \rangle =0$. The two are related by the fluctuation-dissipation relation, $\langle \xi ({\bf x},t) \xi ({\bf x'},t')\rangle=2\eta\theta\delta({\bf x}-{\bf x'})\delta(t-t')$. $\theta\equiv T/a$ is the dimensionless temperature and we take $k_B=\hbar=1$. 

\begin{figure}[b]
%[htbp]
\includegraphics[width=\linewidth]{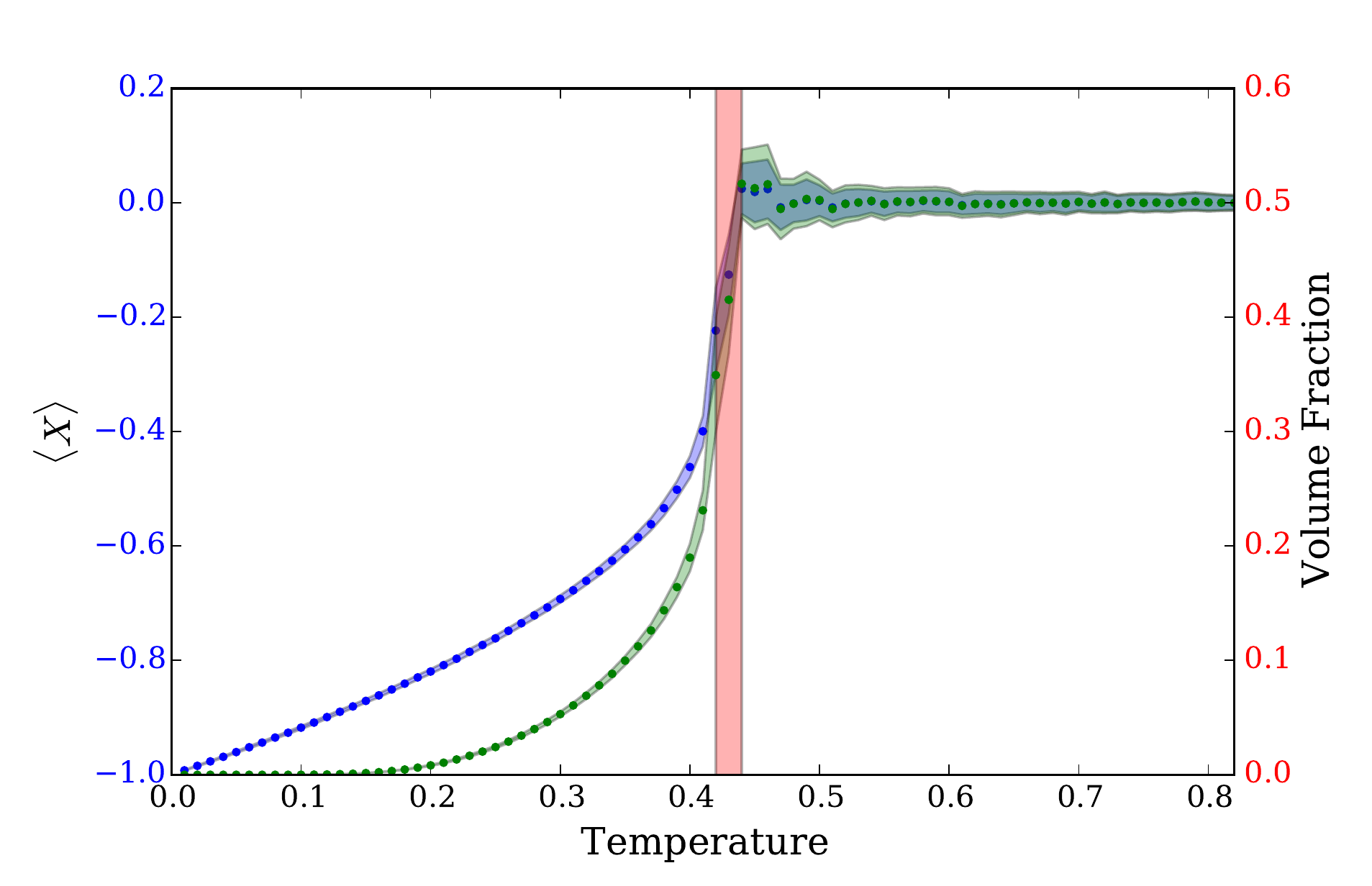}
\caption{(Color online.) The order parameter $\langle X\rangle$ [top (blue) line] and the volume fraction ($p_V$) occupied by the $X > 0$ phase [bottom (green) line] vs. temperature. The shaded regions correspond to 1$\sigma$ deviations from the mean. Within the accuracy of our simulation, the critical temperature is $\theta_c \simeq 0.43 \pm .01$, marked by the vertical band.}
\label{PhivsT}
\end{figure}

We use a staggered leapfrog method and periodic boundary conditions to implement the simulation in a square lattice of size $L$ and spacing $h$. We used $L=100$ and $h=0.25$. Simulations with larger values of $L$ produce essentially similar results, apart from typical finite-size scaling effects \cite{Goldenfeld}. Different values of $h$ can be renormalized with the addition of proper counter terms, as has been discussed in Refs. \cite{Parisi} and \cite{BG}. Since all we need here is to simulate a phase transition with enough accuracy, we leave such technical issues of lattice implementation of effective field theories aside. We follow Ref. \cite{BG} for the implementation of the stochastic dynamics, so that the noise is drawn from a unit Gaussian scaled by a temperature dependent standard deviation, $\xi=\sqrt{2\eta\theta/\Delta t h^2}$. To satisfy the Courant condition for stable evolution we used a time-spacing of $\Delta t =h/4$. The discrete Laplacian is implemented via a maximally rotationally invariant convolution kernel \cite{Lindeberg} with error of $O(h^2)$.

We start the field $X$ at the minimum at $X=-1$ and the bath at low temperature, $\theta=0.01$. We wait until the field equilibrates, checked using the equipartition theorem: in equilibrium, the average kinetic energy per degree of freedom of the lattice field, $\langle\dot X_{ij}^2/2\rangle$, is $\langle \dot X^2_{ij}\rangle = \theta$. Once the field is thermalized, which typically takes about $1,000$ time steps, we use ergodicity to take $200$ readings separated by 50 time steps each to construct an ensemble average. We then increase the temperature in increments of $0.01$ and repeat the entire procedure until we cover the interval $\theta\in[0.01,0.83]$.  

We then obtain the ensemble-averaged $\langle X\rangle$ vs. $\theta$. The coupling to the bath induces temperature-dependent fluctuations which, away from the critical point, can be described by an effective temperature-dependent potential, as in the Hartree approximation \cite{Hartree}. The critical point occurs as $\langle X\rangle\rightarrow 0$, when the $\mathbb{Z}_2$ symmetry is restored. 

\begin{figure*}[t]
\includegraphics[width=\linewidth]{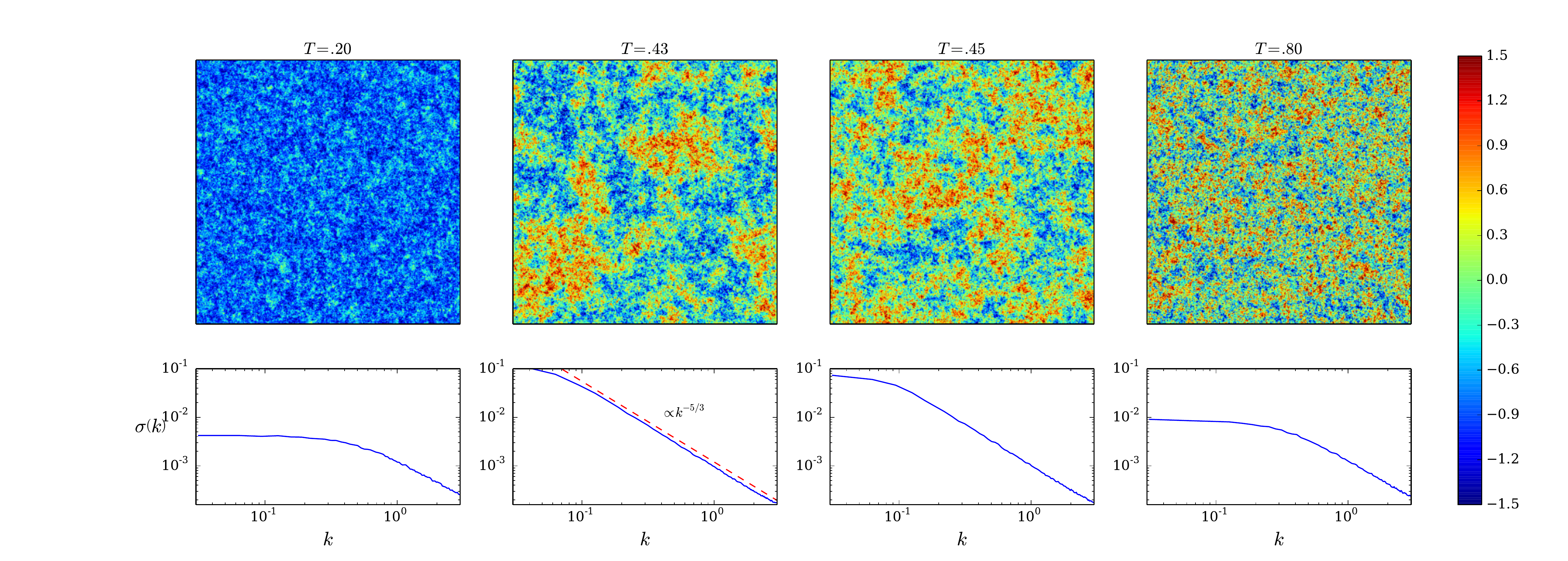}
\caption{(Color online.) Top Row: Snapshots of the equilibrium field $X(x,y)$ for different temperatures. The bar on the right denotes the field magnitude. Bottom Row: The corresponding mode distribution of the CE density. Far from the critical temperature, the CE density is scale-invariant for low $k$, while close to the critical temperature flow into IR produces scaling suggestive of turbulent behavior, with  power-law $k^{-5/3}$ as in Kolmogorov turbulence \cite{Kolmogorov}.}
\label{ModePopulation}
\end{figure*}

In Fig. \ref{PhivsT} we plot the results. The top (blue) line is $\langle X\rangle$, while the bottom (green) line is the ensemble-averaged fraction of the volume occupied by $X > 0$ ($p_V$). Symmetry restoration corresponds to this fraction approaching 0.5.
Shadowed regions correspond to $1\sigma$ deviation from the mean. Within the accuracy of our simulation, the critical temperature is $\theta_c \simeq 0.43 \pm .01$. In the top row of Fig. \ref{ModePopulation} we show the field at different temperatures, including at $\sim T_c$, where large-size fluctuating domains are apparent, indicative of the divergent correlation length at $\theta_c$. 

\section{Configurational Entropy of the Critical Point}
Consider the set of square-integrable bounded periodic functions with period $L$ in $d$ spatial dimensions, $f(\textbf{x}) \in L^2({\mathbb{ R}^d})$, and their Fourier series decomposition, $f({\bf x})=\sum_{{\bf k}_{\bf n}}F({\bf k}_{\bf n})e^{i{{\bf k}}_{\bf n}\cdot {\bf x}}$, with ${\bf k}_{\bf n}=2\pi(n_1/L,\dots, n_d/L)$, and $n_i$ integers. Now define the modal fraction $f_{\bf {k}_{\bf n}}=|F({\bf k}_{\bf n})|^2/ \sum |F(\bf{k}_{\bf n})|^2$. (For details and the extension to nonperiodic functions see \cite{GS1}.) The configurational entropy for the function 
$f(\bf x)$, $S_C[f]$, is defined as 
\begin{equation}
\label{CE}
S_C[f] = - \sum f_{\bf {k}_{\bf n}}\ln[f_{\bf {k}_{\bf n}}].
\end{equation}
The quantity  $\sigma({\bf k_n})\equiv -f_{\bf {k}_{\bf n}}\ln[f_{\bf {k}_{\bf n}}]$ gives the relative entropic contribution of mode $\bf {k}_{\bf n}$. In the spirit of Shannon's information entropy \cite{Shannon}, $S_C[f]$ gives an informational measure of the relative weights of different $k$-modes composing the configuration: it is maximized when all $N$ modes carry the same weight, the mode equipartition limit, $f_{\bf k_n}=1/N$ for any $k_n$, with $S_C[f]=\ln N$. If only a single mode is present, $S_C[f]=0$. For the lattice used here, with $N = 400^2$ points, the maximum entropy is $S_C^{\rm max} = 11.98$.

\begin{figure}[htbp]
\includegraphics[width=\linewidth]{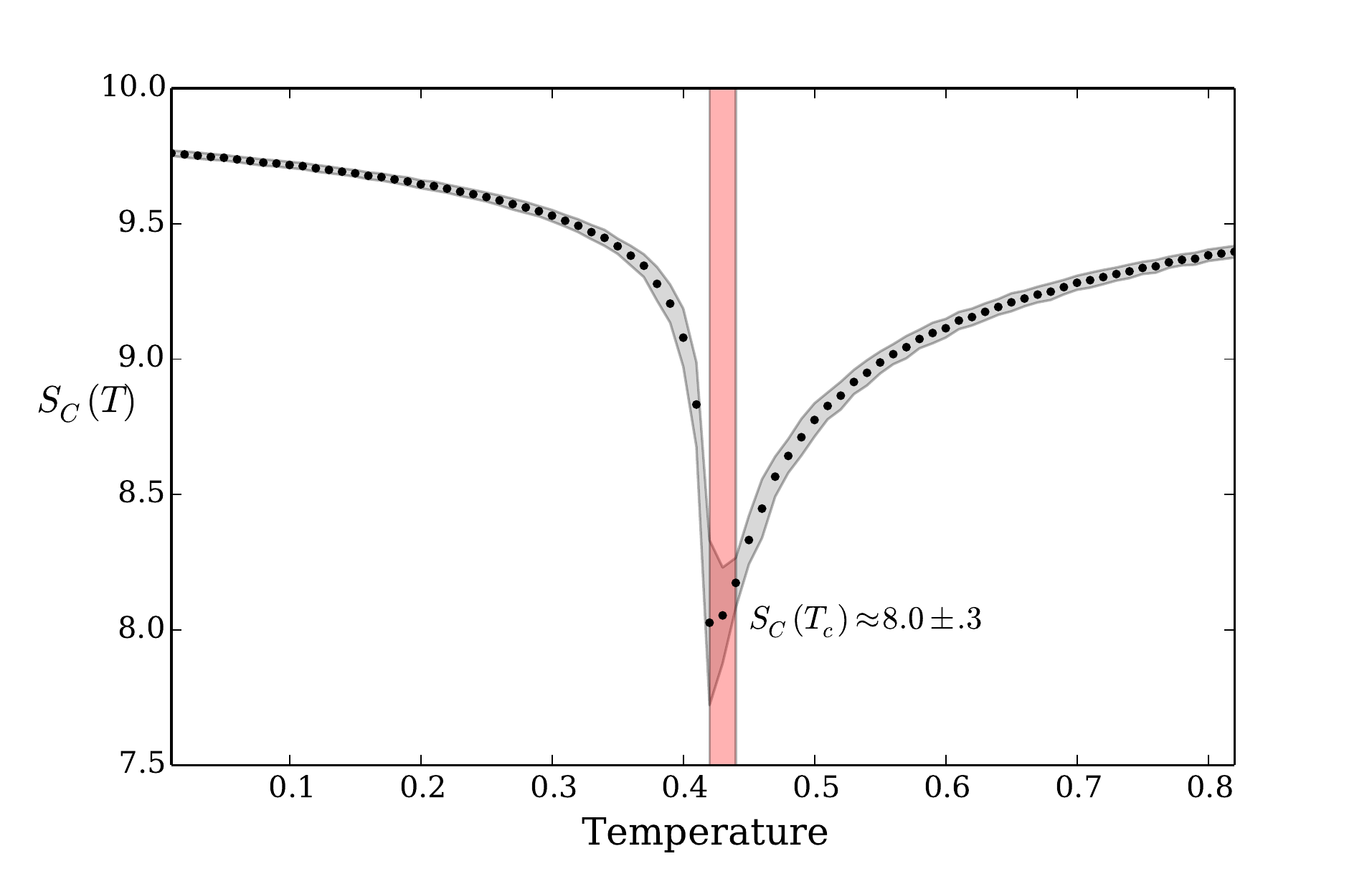}
\caption{(Color online.) Configurational entropy vs. $\theta$ for the ensemble-averaged field $\langle X\rangle$. The minimum at $\theta_c$ is apparent. The lighter shade of grey corresponds to a $1\sigma$ deviation from the mean.}
\label{CEPlot}
\end{figure}

Consider what a field profile looks like for the above examples. Plane waves in momentum space have equally distributed modal fractions, and their position space representations are highly localized. Conversely, singular modes in momentum space have plane wave representations in position space which are maximally delocalized. Localized distributions in position space maximize CE (many momentum modes contribute), while delocalized distributions minimize it. $S_C[f]$ is, in a sense, an entropy of shape, an informational measure of the complexity of a given spatial profile in terms of its momentum modes. The lower $S_C[f]$, the less information (in terms of contributing momentum modes) is needed to characterize the shape.
In the context of phase transitions, we should expect $S_C[f]$ to vary  at different temperatures, as different modes become active. In particular, given that near $T_c$ the average field distribution is dominated by a few long-wavelength modes, we should expect a sharp decrease in the CE.

We use Eq. \ref{CE} to compute the configuration entropy of $\phi(x,y)$. The discrete Fourier transform is 
\begin{equation}
{\bf \Phi}(k_x,k_y) = \mathcal{N}\sum_{n_x=0}\sum_{n_y=0}\phi(n_x h,n_y h)e^{-ih(n_x k_x +n_y k_y)}, 
\end{equation}
\noindent
where $k_{x(y)}=2\pi m_{x(y)}/{L},$ and $m_{x(y)}$ are integers.
We display the CE density as a function of mode magnitude $|k|$ in the bottom row of Fig. \ref{ModePopulation}. For temperatures far from the critical value, we find a scale-invariant CE density for long wavelength modes, and a tapering distribution for short wavelengths. As the critical temperature is approached, the field configurations are dominated by large wavelength modes, and the CE density flows toward lower values of $|k|$, disrupting the scale invariance of the spectrum. This is expected from the diverging correlation length that characterizes the critical point. We find a scaling behavior near $T_c$ with slope $k^{-5/3}$, the same as Kolmogorov turbulence in fluids \cite{Kolmogorov}. Following the analogy, we call this critical scaling {\it information-entropic turbulence}, here with modes flowing from the UV into the IR.

The flow of CE-density into IR modes illustrates how lower-$k$ modes occupy a progressively larger volume of momentum space as the critical point is approached. Hence the sharp decline in CE near $T_c$. In fact, a simple mean-field estimate predicts that the CE will approach zero at $T_c$ in the infinite-volume limit: using that in mean-field theory $\xi(T) \sim |T-T_c|^{-1/2}$ and considering the dominant fluctuations far away from $T_c$ as being Gaussians with radius $\xi(T)$, we can use the continuum limit to compute the CE of a single Gaussian in $d=2$ as \cite{GS1}: $S_C[T] = 2\pi/\xi^2 = 4\pi (1-T/T_c)$. Although even far away from $T_c$ we shouldn't expect this simple approximation to match the behavior of Fig. \ref{CEPlot}, the general trend is for CE to vanish at $T_c$ in the infinite volume limit. In a finite lattice of length $L$, there is going to be a minimum value for CE given by the largest average fluctuation within the lattice. Indeed, in Fig. \ref{CEPlot} we see that the critical point is characterized by a minimum of the CE at $S_C(\theta_c)\simeq 0.80\pm 0.30$. We found that as a function of $|\theta-\theta_c|$ the CE drops super-exponentially as criticality is approached from below, whereas from above we obtain the approximate scaling behavior $S_C[\phi](\theta)\propto|\theta-\theta_c|^{-\frac{1}{4}}$.

\begin{figure}[htbp]
\includegraphics[width=\linewidth]{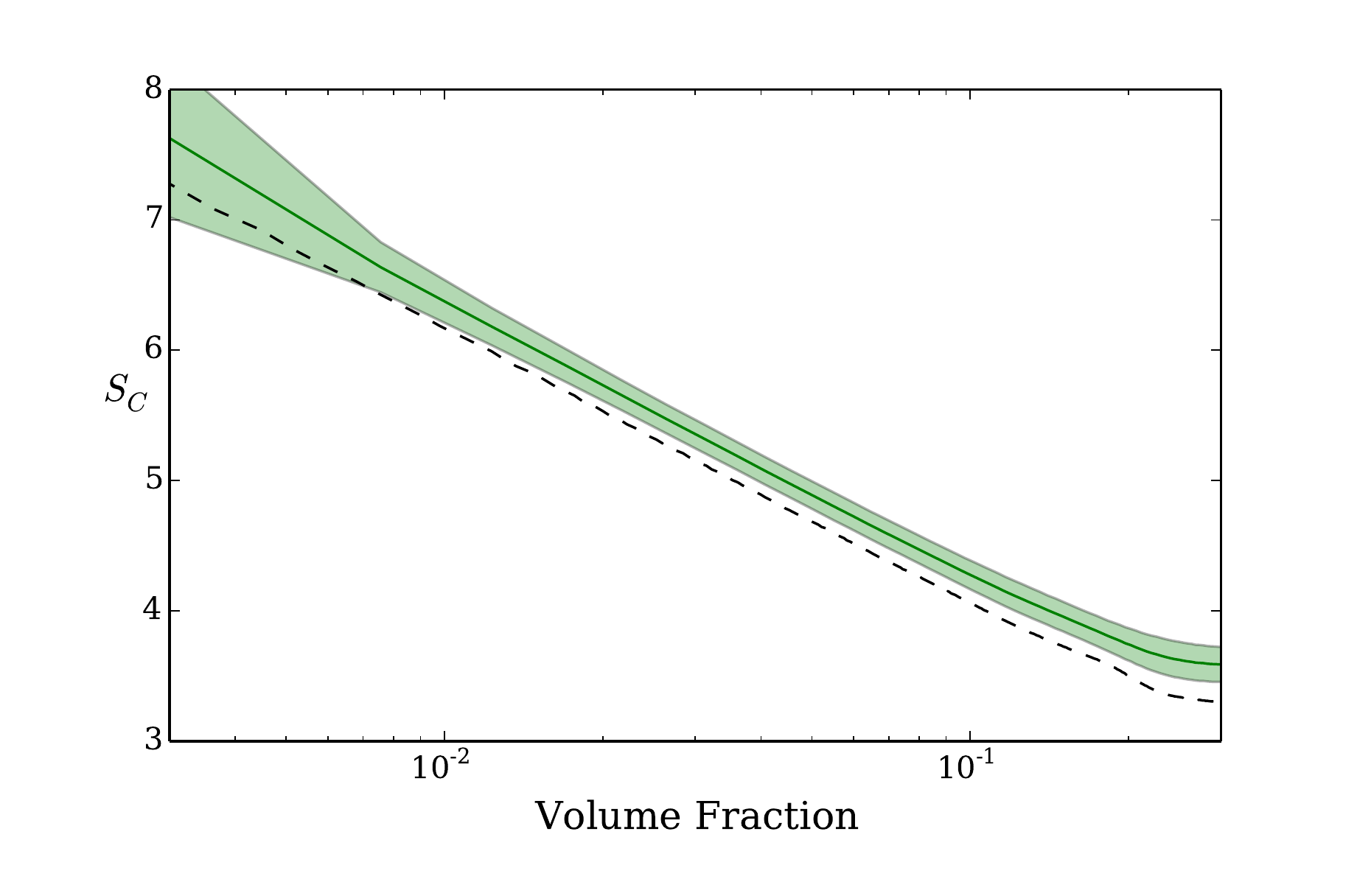}
\caption{(Color online.) Configurational entropy vs. the log of volume fraction occupied by a single fluctuation with $\tanh$ profile defined in Eqs. \ref{TanhBubble} and \ref{Random}. The shadowed area corresponds to 1$\sigma$ deviation from mean. The dashed line corresponds to a symmetric bubble.}
\label{TanhBubPlot}
\end{figure}

\begin{figure}[htbp]
\includegraphics[width=\linewidth]{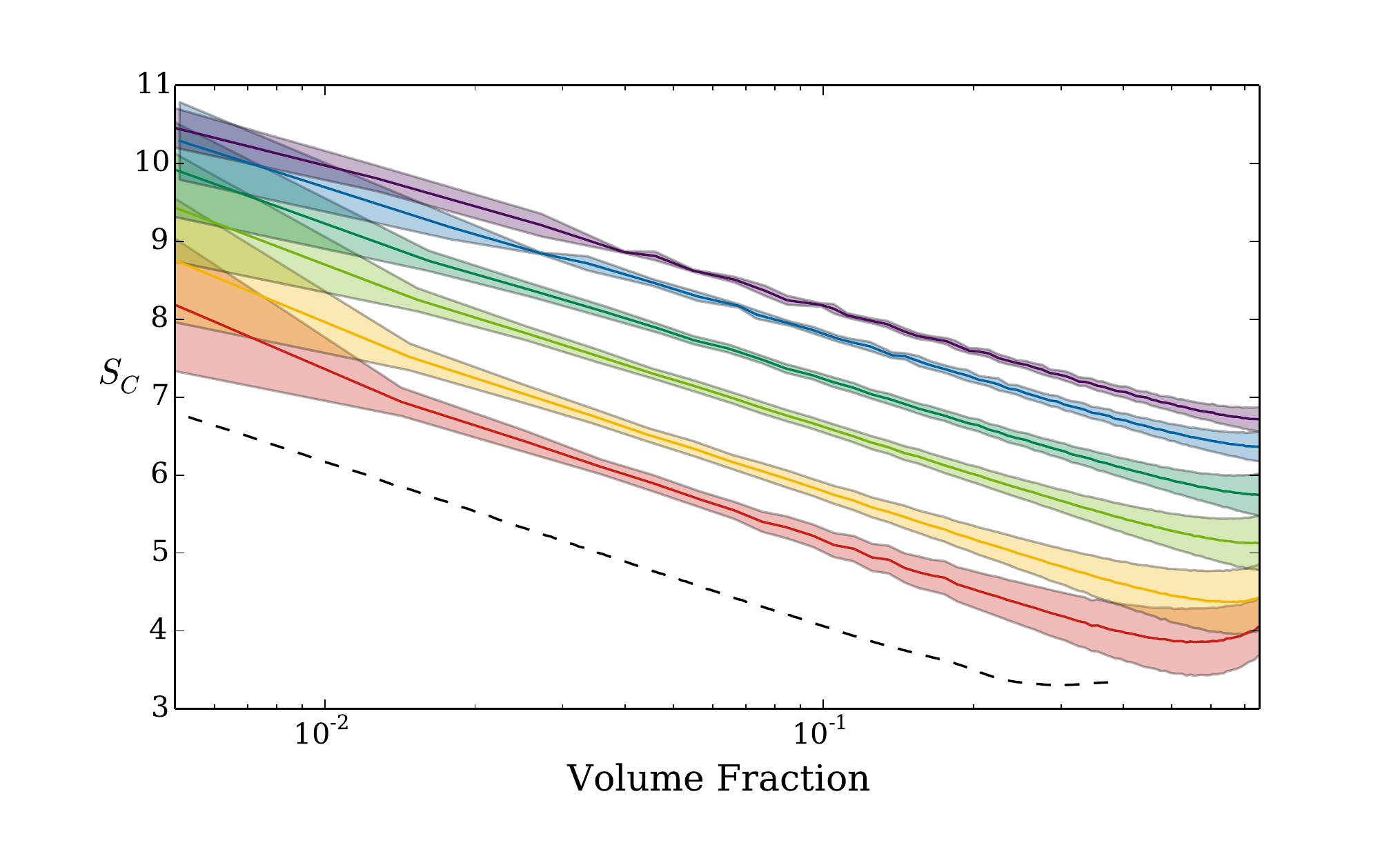}
\caption{(Color online.) Configurational entropy vs. the log of volume fraction for ensembles of bubbles placed at random positions in lattice of side $L=100$. From bottom up, the number of bubbles is 5, 10, 25, 50, 100, and 150. The shadowed areas corresponds to 1$\sigma$ deviation from the mean. The dashed line corresponds to a single bubble with tanh profile defined in Eqs. \ref{TanhBubble} and \ref{Random}.}
\label{Rainbow}
\end{figure}

For a more realistic estimate of the minimum value of CE for a finite lattice near criticality, we model a large fluctuation as a domain of radius $R$ and a kink-like functional profile 
\begin{equation}
\label{TanhBubble}
\phi(r,\varphi) = \frac{1}{2}\left [1 + {\rm tanh}\left (\frac{R - a(\varphi)r}{d}\right )\right ],
\end{equation}
\noindent
where $d$ measures the thickness of the domain wall in units of lattice length $L$ and $a$ is a random perturbation defined as
\begin{equation}
\label{Random}
a(\varphi) = 1 + \sum_{n=3}^{10} \frac{\alpha_n}{n}\cos\left (n\varphi + \beta_n \right ),
\end{equation}
\noindent
with $\alpha_n$ and $\beta_n$ being uniformly-distributed random numbers defined in the intervals $(0,1)$ and  $(0,2\pi)$, respectively. We took $d=0.01L$ so that the domain wall thickness matches the zero-temperature correlation length. Eqs. \ref{TanhBubble} and \ref{Random} define a fluctuation interpolating between $\phi=0$ and near $\phi=1$. In Fig. \ref{TanhBubPlot} we show the results for an ensemble of such  fluctuations occupying different volume fractions. Note how the symmetric fluctuation (dashed line, obtained by setting $a(\varphi) = 1$ in Eq. \ref{TanhBubble}) has lower CE than the asymmetric fluctuations, as one should expect given that CE measures spatial complexity.
Fluctuations with $R\geq L/4$ (volume fraction $\gtrsim 0.2$) begin to have boundary issues due to the tail of the configuration. If we thus take a fluctuation with $R/L= 0.25$ to represent large fluctuations near $T_c$, from Fig. \ref{TanhBubPlot} its ensemble-averaged CE is ${\rm CE} \sim 3.72\pm 0.13$. Of course, a single bubble doesn't match the complexity of the percolating behavior at $T_c$. To simulate the percolating transition, we placed different numbers of bubbles randomly on the lattice with initial radius $R/L=0.001$. We then let their radius $R$ grow to $R/L=0.1$, measuring the volume fraction ($p_V$) they occupy, while computing the CE as their radii increase.
In Fig. \ref{Rainbow} we show the CE as a function of $p_V$. From bottom up, the number of bubbles is 5, 10, 25, 50, 100, and 150. The dashed line corresponds to a symmetric bubble. As the number of bubbles increases, the CE at $p_V=0.5$ approaches the numerical value at $T_c$ (see Fig. \ref{CEPlot}). The results show a simple behavior, $S_C(N,p_V) \sim B(N)\log(p_V/L^2)$, with $B(N)$ having a weak $N$ dependence.
%The results can be fitted by a simple function, $S_C(N,p_V) \simeq A(N) + B(N)\log(p_V/L^2)$, with
%$A(N) \simeq 1.8[1+ 0.5\log(N)]$ and $B(N) \simeq 0.85$, with a weak $N$-dependence.

We have presented a new diagnostic tool to study critical phenomena based on the configurational entropy (CE). We have shown how criticality implies in a sharp minimum of CE, and identified a transition from scale-free to scaling behavior at criticality, with the same power law as Kolmogorov's turbulence. We introduced a percolating-bubble model to describe our numerical results. We are currently exploring the notion of informational turbulence and its relation to more complex percolation models with varying bubble sizes and hope to report on our results soon.

\acknowledgements The authors thank Adam Frank for useful discussions. MG and DS were supported in part by a Department of Energy grant DE-SC0010386. MG also
acknowledges support from the John Templeton Foundation grant no. 48038.

\end{document}